\documentclass[Journal]{IEEEtran}
\IEEEoverridecommandlockouts

\usepackage{cite}
\usepackage{amsmath,amssymb,amsfonts}
\usepackage{algorithmic}
\usepackage{graphicx}
\usepackage{textcomp}
\usepackage{xcolor}
\newcommand{\added}[1]{#1}

\def\BibTeX{{\rm B\kern-.05em{\sc i\kern-.025em b}\kern-.08em
    T\kern-.1667em\lower.7ex\hbox{E}\kern-.125emX}}
\begin{document}

\title{Automatic Modulation Classification via \\ Green Machine Learning }
\author{
    \IEEEauthorblockN{Chee-An Yu$^{1}$}, \IEEEauthorblockN{Young-Kai Chen$^{2}$}, \IEEEauthorblockN{C.-C. Jay Kuo$^{1}$}    \\
    \IEEEauthorblockA{\textit{$^{1}$ University of Southern California, Los Angeles, USA}} \\
    \IEEEauthorblockA{\textit{$^{2}$ Coherent Corp. Santa Clara, CA, USA}}\\
}


\maketitle

\begin{abstract}
In this work, we propose an interpretable, robust, and lightweight machine learning method for automatic modulation classification (AMC) under dynamic and noisy channel conditions. It is called green automatic modulation classification (GAMC) and targets edge artificial intelligence (AI) with low computational complexity and a small model size. GAMC operates in four stages. First, raw received I/Q signals are transformed into multi-domain representations, including constellation diagrams and spatio-temporal graphs. Second, we extract a comprehensive set of statistical and topological features from time-series signals, constellation diagrams, and graphs. Third, a supervised feature learning process leverages label guidance to project high-dimensional features into robust, discriminative low-dimensional ones. Finally, a context-aware Signal-to-Noise Ratio (SNR) soft routing mechanism ensembles predictions from downstream classifiers. Experimental results show that GAMC effectively mitigates domain shifts caused by high noise. It strikes a good balance between accuracy and efficiency, reducing the number of model parameters by $50\%$, operating at $3\%$ to $42\%$ of the computational cost of lightweight deep learning models, and maintaining higher accuracy in various SNRs.
 
\end{abstract}


\section{Introduction}
\begin{figure}[htbp]
\centering
\includegraphics[width=0.9\columnwidth]{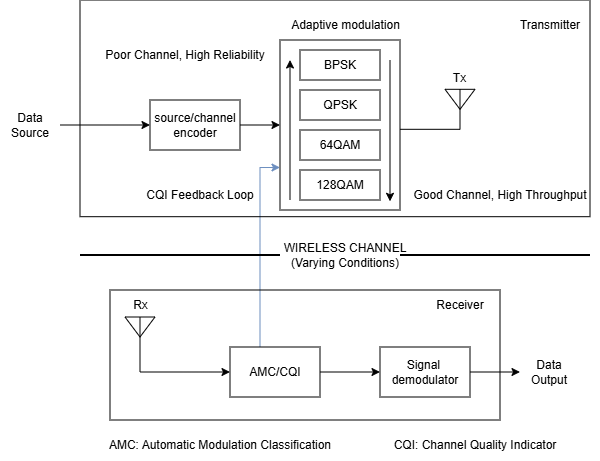}
\caption{The block diagram of an adaptive modulation communication system.}\label{overview of AMC}
\end{figure}

Automatic modulation classification (AMC) is a fundamental technology in modern wireless communications, serving as a critical step for a receiver to demodulate and decode a transmitted signal without knowing the transmitter's settings in the adaptive modulation context, as shown in Fig. \ref{overview of AMC}. In an increasingly congested Radio Frequency (RF) spectrum, AMC is the backbone of intelligent communication systems, enabling dynamic spectrum access, interference identification, and spectrum monitoring. Its applications include civilian and military domains, playing an indispensable role in cognitive radio networks, 5G/6G adaptive modulation schemes, electronic warfare, and threat analysis. As edge devices and Internet of Things (IoT) networks proliferate, the demand for fast, accurate, and autonomous modulation recognition has never been higher.

Historically, AMC algorithms have been divided into two primary categories: likelihood-based and feature-based methods. Likelihood-based methods treat modulation recognition as a composite hypothesis-testing problem, achieving mathematically optimal classification accuracy \cite{zheng2025recent,wei2019automatic, wei2000maximum, chang2021multitask, sills1999maximum,ge2021accuracy,abu2018automatic}. However, they suffer from prohibitive computational complexity and are highly sensitive to channel impairments such as phase offsets and timing synchronization errors. Feature-based methods, conversely, extract human-crafted statistical features, such as higher-order cumulants, cyclic moments, and spectral characteristics—and feed them into traditional classifiers\cite{wu2008novel,ramkumar2009automatic,zhang2019automatic,wang2017graphic,zhu2014genetic, wang2010fast}. Although feature-based methods are computationally efficient and often physically interpretable, their performance often degrades sharply in severe noise environments. Recently, Deep Learning (DL) methods that use Convolutional Neural Networks (CNNs), Recurrent Neural Networks (RNNs), and Transformers have dominated the field~\cite{o2018over,zheng2023mobilerat,ahmadi2025enhancing}. By operating directly on raw in-phase and quadrature (I/Q) samples or time-frequency representations, these models achieve state-of-the-art accuracy and automatically map complex, nonlinear noise manifolds. However, this performance comes at a steep cost. DL models demand massive computational resources and vast amounts of training data, and are notoriously difficult to deploy on power-constrained edge receivers.

Beyond computational and power constraints, the most critical drawback of DL in AMC is its "black-box" nature, which lacks interpretability. In a mission-critical communication infrastructure, an opaque model that prevents diagnosis of prediction failures is a severe limitation. To address this, we propose a lightweight, explainable, and sustainable learning pipeline for AMC, called green automatic modulation classification (GAMC). \added{Although our earlier work in \cite{yu2025gamc} investigated the green AMC problem, the current work is not a straightforward extension of \cite{yu2025gamc}. Instead, we introduce a new set of techniques that are substantially different and novel.} GAMC bridges the gap between traditional feature-based methods and modern end-to-end deep learning. By utilizing SNR-aware tree-based ensembles and supervised linear projections, GAMC autonomously distills complex spatio-temporal interactions while maintaining a fully transparent pipeline. Mathematical transformations, feature selection mechanisms, and routing decisions can be explicitly tracked and physically interpreted. The result is a highly efficient, edge-friendly architecture that provides robust accuracy without sacrificing sustainability or human insight.

To achieve this transparent and lightweight paradigm, we introduce the green learning framework that systematically transforms raw signals into highly discriminative predictions without relying on opaque deep neural networks. The contributions of this work are summarized as follows:
\begin{itemize}
    \item First, to mitigate the severe degradation of signal structures in high-noise environments, we introduce a spatio-temporal graph modeling approach that explicitly captures the topological features of the constellation points, showing improved robustness under low Signal-to-Noise Ratio (SNR) conditions.
    \item Second, we devise a linear feature projection step—acting as a supervised feature learning mechanism—to generate low-dimensional, highly discriminative features downstream tree-based classifiers to maximize their use.
    \item Third, we propose an SNR-aware soft-routing Mixture of Experts (MoE) architecture to stabilize the volatile decision boundaries inherent in extreme SNR variations. By seamlessly interpolating between specialized expert models based on real-time channel quality, this mechanism prevents severe performance degradation typically caused by domain shifts in the feature space.
\end{itemize}
The above innovations establish a robust, interpretable, and computationally sustainable solution to modern challenges in AMC.

\section{Review of Related Work}

This section reviews likelihood-, feature-, and learning-based AMC methods.

\subsection{Likelihood-based Methods}

Likelihood-based methods treat AMC as a decision-theoretic problem \cite{wei2019automatic, wei2000maximum, chang2021multitask, sills1999maximum}. They estimate the likelihood of each modulation hypothesis from the observed signal and classify by maximizing the likelihood. Although they offer optimal classifiers in the Bayesian sense, likelihood-based methods incur high computational costs because they must evaluate likelihoods across multiple candidate modulations \cite{ge2021accuracy}. In addition, they require extensive prior knowledge of signal parameters—such as carrier frequency, phase, and symbol rate—which significantly reduces their robustness in practical scenarios. To address these vulnerabilities, \cite{abu2018automatic} proposed a hybrid scheme that integrates statistical moments. By combining invariant moment-based features with the LB approach, this method effectively mitigates the system's sensitivity to unknown channel effects. Overall, while LB methods have a solid foundation in statistical decision theory and well-understood error bounds, their high computational burden and limited robustness limit their real-world applicability. These constraints have driven the field to explore more flexible, practical Feature-Based (FB) methods.

\subsection{Feature-based Methods}

Rather than maximizing the likelihood of the raw waveform, feature-based methods extract discriminative signal signatures and use a classifier to determine the modulation type. They generally operate with significantly lower computational complexity and require far less prior information than likelihood-based methods. Although feature-based methods are suboptimal in the strict Bayesian sense \cite{ghasemzadeh2018performance}, their flexibility and efficiency are attractive for real-world deployments. A diverse array of features has been proposed for AMC. Common ones include higher-order statistics \cite{wu2008novel}, cyclostationary features \cite{ramkumar2009automatic}, time-frequency distributions \cite{zhang2019automatic}, constellation diagrams \cite{wang2017graphic}, and cumulative distribution functions \cite{zhu2014genetic, wang2010fast}. They are designed to be invariant to the nuisance parameters in the received signal, such as channel gain and phase offset. Although feature-based methods offer greater robustness and practicality, feature selection and the associated decision rule require manual tuning by domain experts. The labor-intensive feature engineering process has led to a shift toward automatically learning features from data.

\subsection{Learning-based Methods}

With the advancement of deep neural networks (DNNs), learning-based methods leverage strong supervision to learn a mapping from input-to-output pairs. DNNs can learn a representation in the latent space via backpropagation to minimize a supervised loss function. Characteristics of different modulation schemes and their evolution dynamics can be implicitly extracted to minimize intra-class distance and maximize inter-class distance. Representative works are reviewed in the following. 

\subsubsection{CNN and RNN}

The use of DNNs, particularly Convolutional Neural Networks (CNNs) and Recurrent Neural Networks (RNNs), has significantly transformed the AMC problem. Early studies in cognitive radio and software-defined radio highlighted the need for automated, intelligent spectrum sensing, paving the way for learning-based methods. Early models used stacked convolutional autoencoders combined with high-order cumulants to obtain robust features in multipath fading channels \cite{zhang2017modulation}. Shortly after, CNNs, specifically residual networks (ResNet), demonstrated the ability to extract spatial and structural features from in-phase and quadrature (I/Q) radio signals over-the-air \cite{liu2017deep, o2018over}. To capture inherent temporal dependencies of sequential radio bursts, researchers subsequently introduced RNN variants, such as Long Short-Term Memory (LSTM) networks and Gated Recurrent Units (GRU). They are often used in low-cost distributed spectrum sensors \cite{rajendran2018deep}. The complementary strengths of CNNs and RNNs led to the development of hybrid architectures, commonly known as the Convolutional Long Short-Term Memory Deep Neural Networks (CLDNNs). Hybrid models have been applied to AMC \cite{pijackova2021radio} and intra-phase modulation radar signal recognition \cite{wei2020intra}. 

Further refinements have focused on improving feature resolution, network robustness, and model efficiency. For instance, recent studies leverage large-kernel multi-scale CNNs combined with Squeeze-and-Excitation (SE) attention mechanisms \cite{wu2022deep}, or employ dilated convolutions to expand the receptive field without exponentially increasing parameter counts \cite{harper2023automatic}. To improve model generalization, techniques such as spectrum-interference-based two-level data augmentation using the Short-Time Fourier Transform (STFT) have been introduced \cite{zheng2021spectrum}, along with a priori regularization methods that increase inter-class distance and reduce intra-class variance \cite{zheng2023dl}. Moving toward recent literature, there is a strong emphasis on extreme efficiency and SNR-awareness. Models like the Threshold Denoise Recurrent Neural Network (TDRNN) achieve high classification accuracy with a remarkably compact footprint \cite {an2025efficient}. Similarly, frameworks such as KADNet integrate SNR-aware deformable convolutions with Kolmogorov-Arnold Networks (KAN) to maintain robust classification under severe noise \cite{wang2026kadnet}. At the same time, complex-valued Hybrid Fusion Deep Neural Networks (HFDNN) have been developed to better exploit the inherently complex nature of baseband signals \cite{tahir2026hfdnn}.

\subsubsection{Transformer}

While CNNs and RNNs excel at extracting local patterns and sequential transitions, they struggle to capture long-range global dependency in extended signal bursts due to restricted receptive fields and vanishing gradients. To overcome this limitation, attention-based Transformers have been developed. Transformers utilize self-attention mechanisms to weigh the significance of all parts of a signal sequence, making them highly adept at handling complex multipath fading channels. Transformer-based DNNs can set new performance benchmarks by globally correlating I/Q samples \cite{hamidi2021mcformer}. Furthermore, specialized variants, such as the Multi-Scale Radio Transformer, employ dual-channel representations to capture fine-grained modulation characteristics at multiple resolutions \cite{zheng2022fine}. As AMC deployment expands into resource-constrained environments, such as drone communications, the demand for lightweight models has grown. MobileRaT exemplifies this shift, offering a highly efficient radio Transformer tailored for unmanned aerial systems \cite{zheng2023mobilerat}.

Furthermore, Transformers are actively used to address data scarcity and noise variation. For example, reconstruction-driven Vision Transformers have been proposed to enhance recognition capabilities using labeled data by learning robust latent representations \cite{ahmadi2025enhancing}. In highly variable noise environments, MoE-AMC effectively employs a Mixture-of-Experts strategy, using a Transformer expert to process low-SNR regimes while routing high-SNR signals to a ResNet expert \cite{gao2023moe}. Additionally, as security becomes paramount, recent studies on adversarial attacks and reliable defenses based on frequency-domain feature enhancement emphasize the need to secure complex Transformers \cite{meng2025adversarial}. The trajectory of recent research highlights the continued convergence towards Attention-enhanced Hybrid Deep Learning-Transformer systems, which seamlessly fuse local CNN feature extraction with global Transformer context modeling \cite{ansari2025attention}.

\subsubsection{GNN}

The vast majority of deep learning approaches in AMC treat radio signals either as grid-like Euclidean data or as flat sequential arrays. However, these traditional representations can obscure the complex, non-Euclidean topological relationships and underlying dynamical system properties inherent in modulated waveforms. To address this representational gap, Graph Neural Networks (GNNs) have recently emerged as a powerful paradigm for modulation classification. The foundational step in applying GNNs to AMC is transforming 1D time-series I/Q data into a structured graph, where nodes represent signal sample points and edges represent the correlation or transition dynamics between them. A prominent example from the recent literature is AvgNet, which leverages an Adaptive Visibility Graph Neural Network \cite{xuan2022avgnet}. By employing a visibility graph algorithm, AvgNet maps time-series data to a complex network, enabling the GNN to capture geometric and structural features that are invisible to standard convolutional filters. This topological perspective is particularly advantageous for distinguishing closely related modulation schemes under severe channel impairments. Beyond visibility graphs, researchers are exploring multi-scale graph constructions combined with LSTM networks or Transformers to mine temporal and spatial graph features simultaneously. Frameworks such as the CNN-Transformer Graph Neural Network (CTGNET) utilize aggregation algorithms to adaptively map signal subsequence matrices, creating highly discriminative feature embeddings \cite {wang2023automatic}. 

\added{\subsubsection{Lightweight and Interpretable AMC}
Beyond pure accuracy improvements, recent AMC research has also begun to emphasize lightweight design, interpretability, and deployment efficiency. In our earlier work, G-AMC~\cite{yu2025gamc}, we investigated a green AMC framework that pursued these objectives through a transparent and computationally efficient pipeline. Although G-AMC followed a different technical route from the present study, it reflects the same broader motivation of developing sustainable and explainable AMC systems for practical wireless environments}.

Despite significant advances, the widespread adoption of deep learning in critical communication systems is fundamentally constrained by its opaque, "black-box" nature \cite{kim2021deep}. This profound lack of interpretability makes diagnosing failures nearly impossible, particularly when models trained on synthetic data encounter unpredictable real-world domain shifts. Because end-to-end architectures obscure their internal decision-making processes, their intermediate feature representations lack physical or theoretical meaning in the context of signal processing. Consequently, the inability to logically trace how a network derives its classifications severely diminishes the reliability, fault-tolerance, and trustworthiness required for practical Automatic Modulation Classification (AMC) deployments.

\begin{figure*}[t]  
\centering
\includegraphics[width=0.85\textwidth]{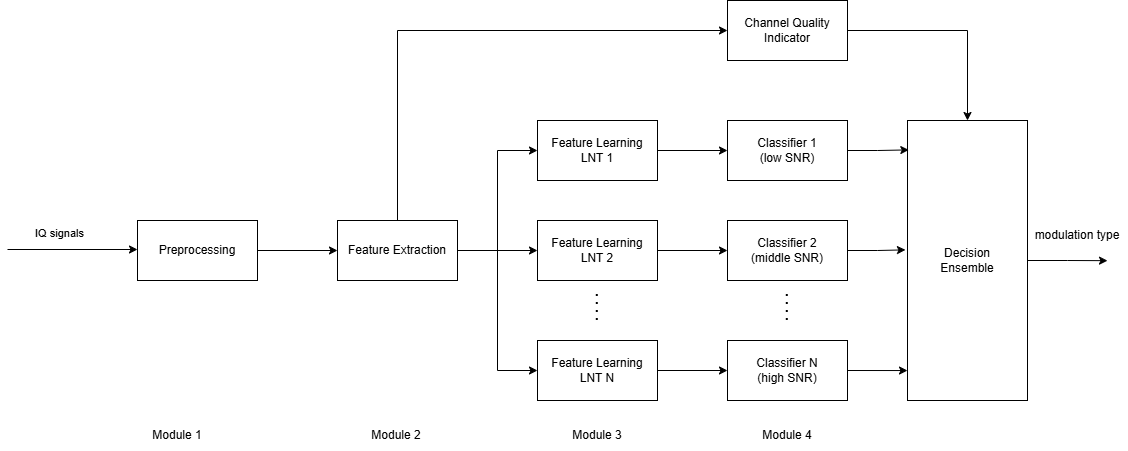}
\caption{Overview of the proposed green automatic modulation classification (GAMC) system.}
\label{overview_pipeline}
\end{figure*}

\section{Green Automatic Modulation Classification}

An overview of the proposed green automatic modulation classification (GAMC) system is illustrated in Fig. \ref{overview_pipeline}. It consists of four modules: 1) preprocessing, 2) feature extraction, 3) feature learning, and 4) SNR-aware mixture of experts (MoEs) classifiers. Module 1 prepares raw data for different representations, such as a constellation map and a spatio-temporal graph. Module 2 is tailored to extract the features from different representations. Module 3 is designed to learn more discriminative features from the previous representations for the downstream classifier. Module 4 consists of XGBoost-based classifiers \cite{chen2016xgboost}, including a channel quality indicator and modulation classifiers.  

\subsection{Preprocessing}

\subsubsection{Constellation Map}

\paragraph{Cartesian Coordinates}
To capture the distinct amplitude and phase characteristics of various modulation schemes, we construct a two-dimensional constellation diagram from the raw signal data. Specifically, each discrete In-Phase ($I$) and Quadrature ($Q$) sample is mapped to a Cartesian coordinate system, with the horizontal axis representing the real $I$ component and the vertical axis representing the imaginary $Q$ component. This spatial representation reveals the underlying geometry of the modulation format, including the distinct circular formations of Phase-Shift Keying (PSK) and the grid-like clusters of Quadrature Amplitude Modulation (QAM). While Cartesian mapping effectively visualizes Additive White Gaussian Noise (AWGN) as symmetric spatial dispersion, its topological structure is highly vulnerable to phase noise and Carrier Frequency Offsets (CFO). These channel impairments induce nonlinear circular rotations across the coordinate plane, potentially scrambling the geometric relationships between neighboring samples.

\paragraph{Polar Coordinates}
To isolate and mitigate these phase-specific impairments, we simultaneously project the $I/Q$ samples into a Polar coordinate manifold. Standard conversion formulas are applied to derive the amplitude ($A = \sqrt{I^2 + Q^2}$) and phase ($\phi = \text{atan2}(Q,I)$) for each sample. This transformation provides powerful physical invariance: it "unwraps" circular PSK constellations into linear, bounded strips. Crucially, CFO and phase noise---which manifest as complex rotations in Cartesian space---are reduced to simple linear translations along the $\phi$-axis in the Polar manifold. Because graph-based topological features are less sensitive to linear translation, the Polar coordinate system provides a fundamentally stable geometric baseline under severe phase distortions and amplitude fading.

\subsubsection{Spatio-Temporal Graph Representation}

\begin{figure*}[htbp]  
\centering
\includegraphics[width=0.85\textwidth]{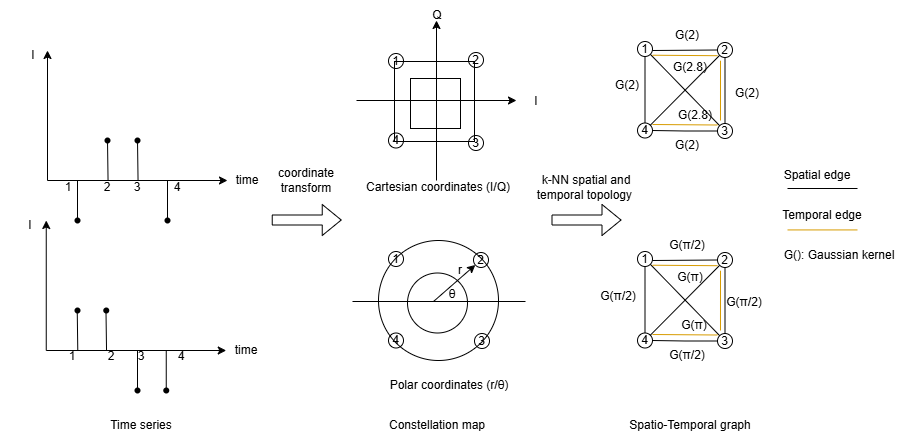}
\caption{The illustration of the signal time series (left), the constellation map (middle), and 
the spatio-temporal graph representation (right).}\label{ST_Graph}
\end{figure*}


To capture the high-order topological relationships within the modulated signals, we construct spatio-temporal graphs modeling interactions across spatial and temporal domains. In this representation, individual signal samples serve as nodes. Connectivity is governed by temporal edges, which preserve chronological order, and by spatial edges established via a k-nearest neighbors (k-NN) algorithm. Rather than relying on a single geometric perspective, we generate complementary dual-manifold graphs by calculating k-NN spatial distances in both coordinate systems. While the physical samples remain identical, altering the distance metric fundamentally changes the resulting graph topology:
\begin{itemize}
\item \textbf{Cartesian Distance:} The Euclidean distance offers a straight-line proximity, $d^2 = \Delta I^2 + \Delta Q^2$. It captures the QAM grid topology and isolates isotropic AWGN clusters natively.
\item \textbf{Polar Distance:} The manifold distance is calculated using independent amplitude and phase differentials, $d^2 = \Delta A^2 + \Delta \phi^2$. This accounts for the circular phase wraparound, enforces strict topological continuity for ring modulations (PSK), and resists phase rotation.
\end{itemize}
These distinct distance calculations provide the downstream model with orthogonal physics-based representations. The Cartesian graph stabilizes QAM grids under AWGN, while the Polar graph stabilizes PSK under phase rotation. To formalize these connections, we pass the respective distance, $\|\mathbf{x}_i - \mathbf{x}_j\|$, through a Gaussian kernel. It transforms unbounded distances into smooth, noise-resistant similarity weights. The spatial adjacency matrix, $\mathbf{A}_s$, and the uniform sequential temporal adjacency matrix, $\mathbf{A}_t$, are constructed as
\begin{equation}
\mathbf{A}_s(i,j) =
\begin{cases}
\exp\left(-\frac{\| \mathbf{x}_i - \mathbf{x}_j \|^2}{\sigma^2}\right), & \text{if } j \in \mathcal{N}_k(i), \\
0, & \text{otherwise}.
\end{cases}
\end{equation}
and
\begin{equation}
\mathbf{A}_t(i,j) =
\begin{cases}
1, & \text{if } |t_i - t_j| = 1, \\
0, & \text{otherwise}.
\end{cases}
\end{equation}
An example is illustrated in Fig. \ref{ST_Graph}.

\subsection{Spatio-Temporal Graph Features and Statistical Features}

\subsubsection{Spatio-Temporal Graph Features}

To enhance robustness under low signal-to-noise ratio (SNR) conditions, we model the received I/Q samples as a spatio-temporal graph. Compared with grid-based or 1D time-array representations, graph topology is inherently more stable against noise perturbations, since structural connectivity is less sensitive to local amplitude fluctuations. Let $\mathbf{A}_s \in \mathbb{R}^{N \times N}$ denote the spatial adjacency matrix constructed from the geometry of the constellation, and $\mathbf{A}_t \in \mathbb{R}^{N \times N}$ denote the temporal adjacency matrix that captures the time-order relationships between the samples. To balance spatial and temporal contributions, we define the spatio-temporal adjacency matrix as
\begin{equation}
\mathbf{A}_{st} = \mathbf{A}_s + \lambda \mathbf{A}_t,
\end{equation}
where $\lambda$ is a hyperparameter that controls the relative importance of temporal connectivity. The degree matrix $\mathbf{D}_{st}$ is a diagonal matrix defined as
\begin{equation}
\mathbf{D}_{st}(i,i) = \sum_{j=1}^{N} \mathbf{A}_{st}(i,j).
\end{equation}
Rather than using the above Laplacian, we compute the symmetric normalized graph Laplacian matrix, given by
\begin{equation}
\mathcal{L}_{st} = \mathbf{I} - \mathbf{D}_{st}^{-1/2} \mathbf{A}_{st} \mathbf{D}_{st}^{-1/2},
\end{equation}
where $\mathbf{I}$ is the identity matrix. We utilized the normalized Laplacian to guarantee scale invariance across varying channel conditions. Because the density of the k-NN spatial graph can fluctuate significantly depending on the SNR level and the specific modulation order (e.g., dense 64-QAM clusters versus sparse BPSK nodes), unnormalized eigenvalues would scale proportionally with the graph's maximum degree. The normalized Laplacian mathematically bounds the entire eigenvalue spectrum to the strict range of $[0, 2]$. This standardization strips away absolute-volume variations, forcing the downstream model to focus solely on the signal's topological structure. According to spectral graph theory, these structural properties are encoded in the eigenvalues of the normalized Laplacian matrix. We perform eigenvalue decomposition
\begin{equation}
\mathcal{L}_{st} = \mathbf{U} \boldsymbol{\Lambda} \mathbf{U}^\top
\end{equation}
where $\boldsymbol{\Lambda} = \mathrm{diag}(\lambda_1, \lambda_2, \dots, \lambda_N)$ contains the eigenvalues sorted in ascending order ($0 \le \lambda_1 \le \lambda_2 \dots \le 2$). The Laplacian spectrum characterizes the smoothness of signals and the consistency of trajectories on the graph. Small eigenvalues correspond to low-frequency structural modes, reflecting global connectivity and macro-clustering behavior, while larger eigenvalues capture high-frequency variations related to local irregularities and noise spikes. Because the spectrum is normalized, these scale-invariant eigenvalues provide a highly consistent structural signature for modulation recognition, effectively isolating distinct constellation geometries and temporal transitions regardless of the raw signal amplitude.

We extract statistical descriptors from the spectrum, including:
\begin{equation}
\text{Spectral entropy} = - \sum_{i=1}^{N} p_i \log p_i,
\quad
p_i = \frac{\lambda_i}{\sum_{j=1}^{N} \lambda_j},
\end{equation}
\begin{equation}
\text{Mean} = \frac{1}{N} \sum_{i=1}^{N} \lambda_i,
\quad
\text{Variance} = \frac{1}{N} \sum_{i=1}^{N} (\lambda_i - \bar{\lambda})^2,
\end{equation}
\begin{equation}
\text{Skewness} = \frac{1}{N} \sum_{i=1}^{N}
\left( \frac{\lambda_i - \bar{\lambda}}{\sigma} \right)^3.
\end{equation}

In addition, the smallest non-zero eigenvalues $(\lambda_1, \lambda_2)$, eigen-gap ratios (e.g., $\lambda_2 / \lambda_1$), and the maximum spectral gap are used to characterize connectivity, bottlenecks, and lattice regularity.

Finally, the neighborhood size parameter $k$ used in graph construction should adapt to SNR conditions. For high SNR, a smaller $k$ preserves fine local geometry and sharp constellation structures. For low SNR, a larger $k$ statistically smooths local density variations and improves robustness against noise-induced fragmentation. In this work, we adopt multiple values of $k$ to account for SNR across a wide range.  

\begin{table*}[t]
\centering
\caption{Spectral graph features and their physical meanings.}
\label{tab:spectral_features}
\begin{tabular}{|l|p{0.7\textwidth}|}
\hline
\textbf{Features} & \textbf{Physical meaning} \\ \hline

$\lambda_1, \lambda_2$ (smallest non-zero eigenvalues) 
& Measure global connectivity and bottlenecks; lower values indicate more clustered or loosely connected constellations. \\ \hline

$\lambda_2 / \lambda_1$ 
& Eigen-gap ratio; amplifies the spectral gap and improves robustness. \\ \hline

Maximum spectral gap 
& Reflects scale separation and regularity of the lattice structure. \\ \hline

Spectral entropy 
& Quantifies structural complexity; higher entropy implies more uniformly distributed modes. \\ \hline

First few eigenvalues 
& Capture fine-scale local geometry and neighborhood density. \\ \hline

Mean / variance/skewness 
& Summarize overall graph smoothness and irregularity. \\ \hline

\end{tabular}
\end{table*}

\subsubsection{Statistical Features}

To complement geometric and graph-based representations, we extract a comprehensive set of statistical features that characterize the amplitude, phase, frequency, and higher-order structure of the received I/Q signal. These features are designed to capture modulation-dependent distributional properties while maintaining robustness under various SNR conditions.

\paragraph{Amplitude-based Statistics}

The amplitude-based features describe the distribution and density structure of the signal magnitude $r[n] = |x[n]|$. Histogram-based features approximate the empirical probability density function (PDF),
\begin{equation}
p(r_k) \approx \frac{1}{N} \sum_{n=1}^{N} \mathbf{1}(r[n] \in \text{bin}_k),
\end{equation}
while cumulative distribution function (CDF) statistics summarize global amplitude structure. Tail-sensitive features quantify the proportion of high-magnitude samples,
\begin{equation}
\eta_{\text{tail}} = \frac{1}{N} \sum_{n=1}^{N} \mathbf{1}(r[n] > \tau),
\end{equation}
which are particularly informative for QAM order discrimination. In addition, two-dimensional I/Q histogram density features preserve absolute amplitude structure and capture lattice regularity in the constellation diagram.

\paragraph{Phase-based Statistics}

Phase-based features analyze the angular component $\theta[n] = \arg(x[n])$. Circular statistics are computed via
\begin{equation}
R = \left| \frac{1}{N} \sum_{n=1}^{N} e^{j\theta[n]} \right|,
\end{equation}
which measures phase concentration and rotational symmetry. Rotational moments of order $k$ are defined as
\begin{equation}
M_k = \left| \frac{1}{N} \sum_{n=1}^{N} x[n]^k \right|,
\end{equation}
and are highly discriminative for $M$-PSK modulations due to cyclic symmetry properties.

To capture temporal phase dynamics, we compute phase differences
\begin{equation}
\Delta\theta[n] = \theta[n] - \theta[n-1],
\end{equation}
and extract both histogram-based and spectral descriptors. These features can be less sensitive to carrier frequency offset (CFO) and effectively distinguish continuous-phase modulations (e.g., FM) from discrete phase transitions (e.g., PSK).

\paragraph{Frequency-based Features}
Frequency-based features are derived from the discrete Fourier transform (DFT),
\begin{equation}
X(f) = \sum_{n=0}^{N-1} x[n] e^{-j2\pi fn/N}.
\end{equation}
Log-magnitude histograms and peak-to-average ratios characterize spectral concentration and bandwidth. Spectral entropy is defined as
\begin{equation}
H = - \sum_f p(f)\log p(f), \quad
p(f) = \frac{|X(f)|^2}{\sum_f |X(f)|^2},
\end{equation}
which quantifies spectral complexity and is useful for distinguishing narrow-band and wide-band modulations.

\paragraph{Higher-Order Statistics}

Higher-order cumulants (HOC) capture the non-Gaussian signal structure while suppressing Gaussian noise contributions. For example, the fourth-order cumulant is given by
\begin{equation}
C_{40} = E[x^4] - 3(E[x^2])^2.
\end{equation}
Such statistics encode constellation geometry and are widely used in classical automatic modulation classification (AMC). In addition, bispectrum-based features,
\begin{equation}
B(f_1,f_2) = E[X(f_1)X(f_2)X^*(f_1+f_2)],
\end{equation}
capture quadratic phase coupling and nonlinear interactions in the frequency domain.

\paragraph{Cyclostationary Features}

Cyclostationary analysis exploits the periodic statistical structure induced by symbol timing and carrier modulation. The cyclic autocorrelation function is defined as
\begin{equation}
R_x^\alpha(\tau) = E\left[x(t)x^*(t-\tau)e^{-j2\pi \alpha t}\right],
\end{equation}
where $\alpha$ denotes the cyclic frequency. These features are highly robust under low SNR and effectively reveal symbol-rate periodicity and modulation-dependent spectral lines.

In summary, the statistical features considered include amplitude, phase, frequency, HOC, and cyclostationary features. They provide complementary descriptors of constellation geometry, temporal dynamics, and periodic structure. This multi-domain statistical representation enhances modulation discriminability and improves robustness under challenging noise conditions.
\subsection{Feature Learning via Linear Projection}\label{AA}

To extract more discriminative representations from the original feature space, a supervised linear projection is used to provide global, complementary decision information to the downstream XGBoost classifier. \cite{wang2024statistics} uses a Least-squares Normal Transform (LNT) to generate new discriminative features with a linear combination of the original features. This effectively bridges the gap between handcrafted representations and learning-based feature synthesis by directly embedding label information into the new features. Specifically, standard tree-based models like XGBoost are restricted to orthogonal axis-aligned splits (e.g., $x_i \leq \theta$). Consequently, they must construct deep, inefficient ``staircase'' (zig-zag) boundaries to approximate a simple diagonal separation. By providing a linearly combined feature, LNT introduces an oblique splitting capability, freeing the downstream XGBoost ensemble to focus its capacity on modeling more complex, nonlinear decision boundaries.

The generation of these linear projection features follows a rigorous three-step process. Let the original feature vector for a given sample be defined as $\mathbf{x} \in \mathbb{R}^D$.

\begin{figure}[htbp]  
\centering
\includegraphics[width=1\columnwidth]{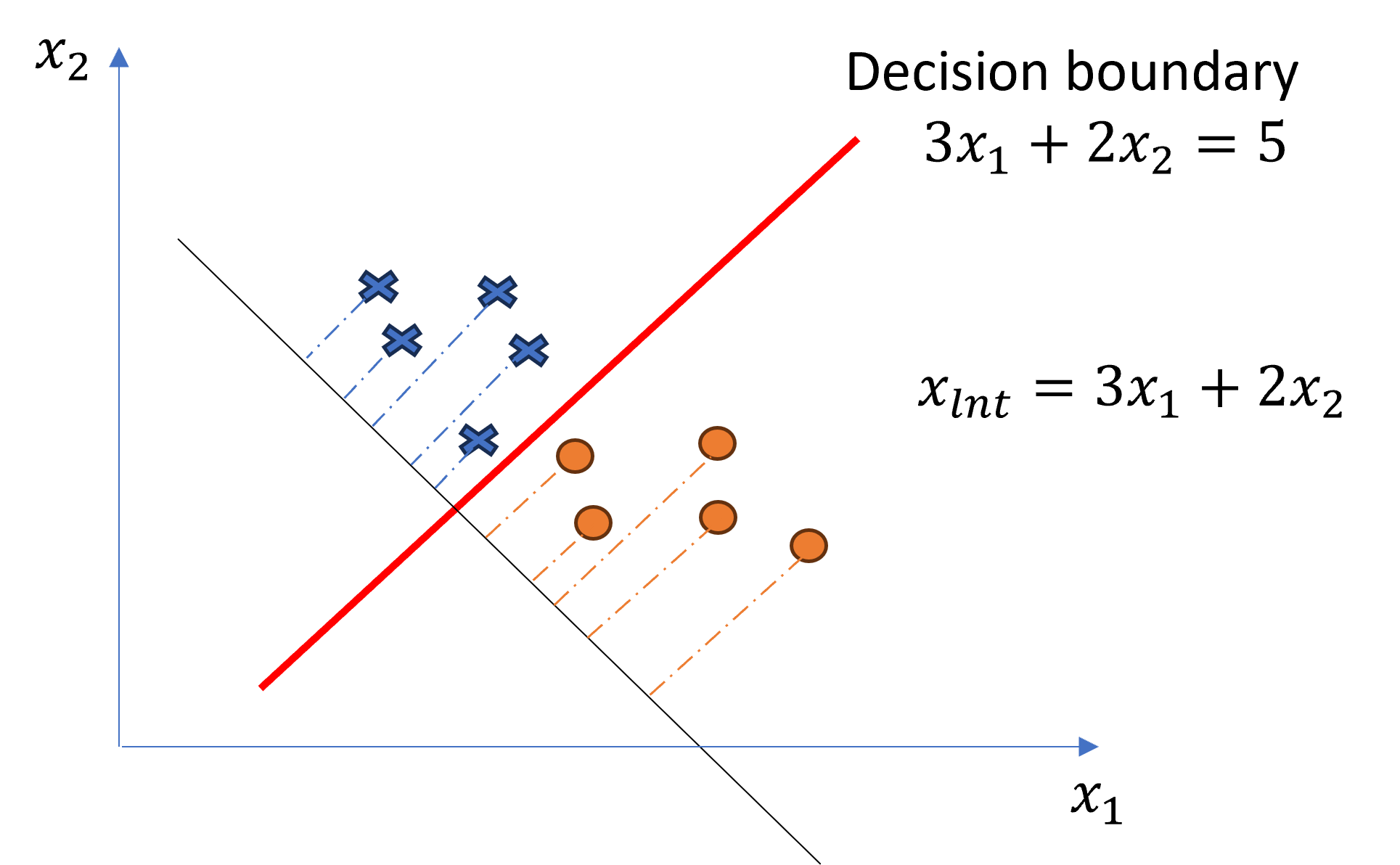}
\caption{Feature learning by linear combination of raw features}
\label{lnt}
\end{figure}

\begin{enumerate}
    \item \textbf{Subspace Feature Selection and Normalization:} First, an auxiliary XGBoost model evaluates the importance of the features to reduce the original $D$-dimensional space. Rather than relying on a single subset, we extract multiple subspaces of features $\mathbf{x}' \in \mathbb{R}^d$ with varying dimensionalities ($d \le D$). This multi-resolution approach balances robustness and accuracy: strictly low-dimensional subspaces act as aggressive noise filters to isolate fundamental topological signatures at low SNRs, while higher-dimensional subspaces retain the fine-grained resolving power needed to distinguish complex modulations at high SNRs. Because the subsequent models are sensitive to scaling, each selected subspace is independently standardized to zero mean and unit variance:
    \begin{equation}
    \mathbf{z} = \frac{\mathbf{x}' - \boldsymbol{\mu}}{\boldsymbol{\sigma}},
    \end{equation}
    where $\boldsymbol{\mu}$ and $\boldsymbol{\sigma}$ denote the mean and standard deviation vectors of the corresponding selected features in the training set.

    \item \textbf{Logistic Regression for Linear Projection:} Second, a one-vs-rest Logistic Regression model is trained using $k$-fold cross-validation to prevent data leakage. For each class $c \in \{1, 2, \dots, C\}$, the model learns a weight vector $\mathbf{w}_c$ and bias $b_c$ to define a linear decision boundary. The input is projected onto this boundary to generate a logit score $l_c$:
    \begin{equation}
    l_c = \mathbf{w}_c^T \mathbf{z} + b_c
    \end{equation}
    An example is illustrated in Fig.~\ref{lnt}.
    \item \textbf{Probability Transformation:} Finally, the projected logits for all $C$ classes are passed through a softmax function to obtain a normalized probability distribution:
    \begin{equation}
    p_c = \frac{\exp(l_c)}{\sum_{j=1}^C \exp(l_j)}
    \end{equation}
    The resulting probability vector $\mathbf{p} = [p_1, p_2, \dots, p_C]^T$ serves as a potent summary of the feature subspace. These probabilities are then concatenated with the original features to enrich the downstream nonlinear XGBoost classifiers.
\end{enumerate}

\subsection{Decision Learning }
\subsubsection{Channel Quality Indicator}
The concatenated spatio-temporal graph features are fed into a Channel Quality Indicator (CQI) to explicitly assess the channel's interference profile. Formulated as a discrete classification task, the CQI categorizes the signal into distinct operational SNR bands (e.g., low, middle, and high). It produces a probability distribution vector—such as $\mathbf{w} = [w_{\text{low}}, w_{\text{mid}}, w_{\text{high}}]$—providing soft assignment weights for each SNR regime, subject to the constraint that their sum equals one.

\subsubsection{Mixture of Experts and Decision Ensemble}
Concurrently, the features are processed by a Mixture of Experts (MoE) architecture comprising three specialized branches. As illustrated in the pipeline overview, each branch pairs a Feature Learning module (LNT) with an expert Classifier. Each pair is trained exclusively on its designated SNR range to optimize for the specific noise distortions within it. To determine the final modulation type, a Decision Ensemble aggregates the outputs of all experts using a soft-routing mechanism. Let $\mathbf{P}_i$ denote the class probability distribution predicted by the $i$-th expert. The final prediction $\mathbf{P}_{\text{ensemble}}$ is the weighted sum of the expert outputs, using the CQI probabilities as weights:
\begin{equation}
\mathbf{P}_{\text{ensemble}} = \sum_{i=1}^{Q} w_i \mathbf{P}_i,
\end{equation}
where $Q$ is the number of experts.
This probabilistic ensembling smoothly interpolates between expert models, mitigating domain shifts and preventing rigid classification errors at noise-floor boundaries.

\section{Experimental Results}

In this section, we verify each module in our Robust GAMC pipeline. The experiments were run on an AMD EPYC 7543 32-core CPU with Ubuntu 22.04 and a NVIDIA RTX A6000 GPU with CUDA 12.04. 

\subsection{Dataset}
We evaluate our proposed method using the widely adopted RadioML 2016.10A dataset \cite{wz93-h307-25}. It is designed to simulate realistic wireless communication scenarios by incorporating various channel impairments, including additive white Gaussian noise (AWGN), multipath fading, frequency offsets, and timing offsets. The dataset comprises 220,000 signal instances across 20 distinct signal-to-noise ratio (SNR) levels, ranging from -20 dB to +18 dB. Each signal observation has a fixed frame length of 128 samples. To ensure a balanced evaluation, the dataset provides exactly 1,000 signals per modulation technique at each SNR level across 11 diverse modulation schemes: AM-DSB, AM-SSB, BPSK, QPSK, 8PSK, QAM16, QAM64, GFSK, CPFSK, PAM4, and WBFM.

\subsection{CQI Classification Accuracy}

\begin{figure}[htbp]  
\centering
\includegraphics[width=1\columnwidth]{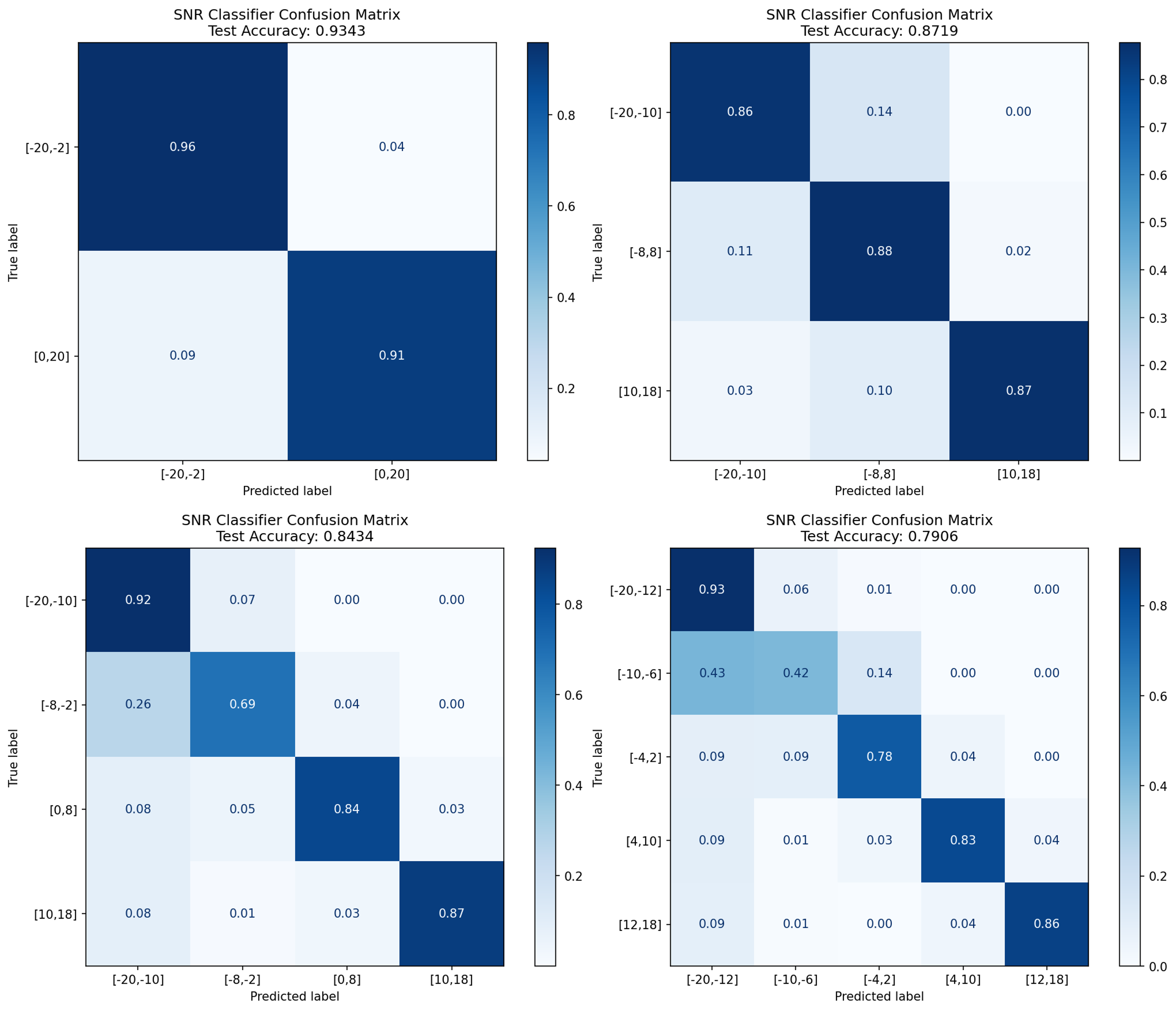}
\caption{Classification Accuracy of Channel Quality Indicator for different SNR bands}
\label{routing_acc}
\end{figure}

To evaluate the impact of routing granularity, the CQI was configured to partition the continuous SNR spectrum into 2, 3, 4, or 5 discrete operational bands. To fulfill the low-power requirements of the green pipeline, the CQI employs a highly lightweight, strictly regularized XGBoost model (detailed in Table \ref{tab:xgboost_params}) with a maximum tree depth of 2 and 200 estimators to prevent overfitting on noisy samples. As Fig.\ref{routing_acc} demonstrates, while increasing granularity naturally reduces strict accuracy (from 93.43\% for 2 bands to 79.06\% for 5 bands), misclassifications are overwhelmingly restricted to adjacent SNR regimes. For example, in the 5-band setup, 43\% of the errors from the [-10, -6] dB band bleed directly into the neighboring [-20, -12] dB band, with zero confusion in distant high-SNR categories. This localized confusion pattern supports the soft-routing Decision Ensemble: because errors remain adjacent, the system smoothly interpolates between neighboring experts without suffering catastrophic misrouting across disparate noise floors.

\begin{table}[htbp]
\centering
\caption{XGBoost Hyperparameters of Channel Quality Indicator Classifier}
\begin{tabular}{|l|c|}
\hline
\textbf{Parameter} & \textbf{Value} \\
\hline
Learning\_rate & 0.1 \\
Max\_depth     & 2 \\
N\_estimators  & 200 \\
colsample\_bytree & 0.8 \\
subsample & 0.8 \\
min\_child\_weight & 1 \\
gamma & 0.1 \\
reg\_alpha & 0.1 \\
reg\_lambda & 0.1 \\
\hline
\end{tabular}
\label{tab:xgboost_params}
\end{table}

\subsection{Performance Evaluation and Ablation Study}

\subsubsection{Spatio-Temporal Graph Feature}

\begin{figure}[htbp]  
\centering
\includegraphics[width=1\columnwidth]{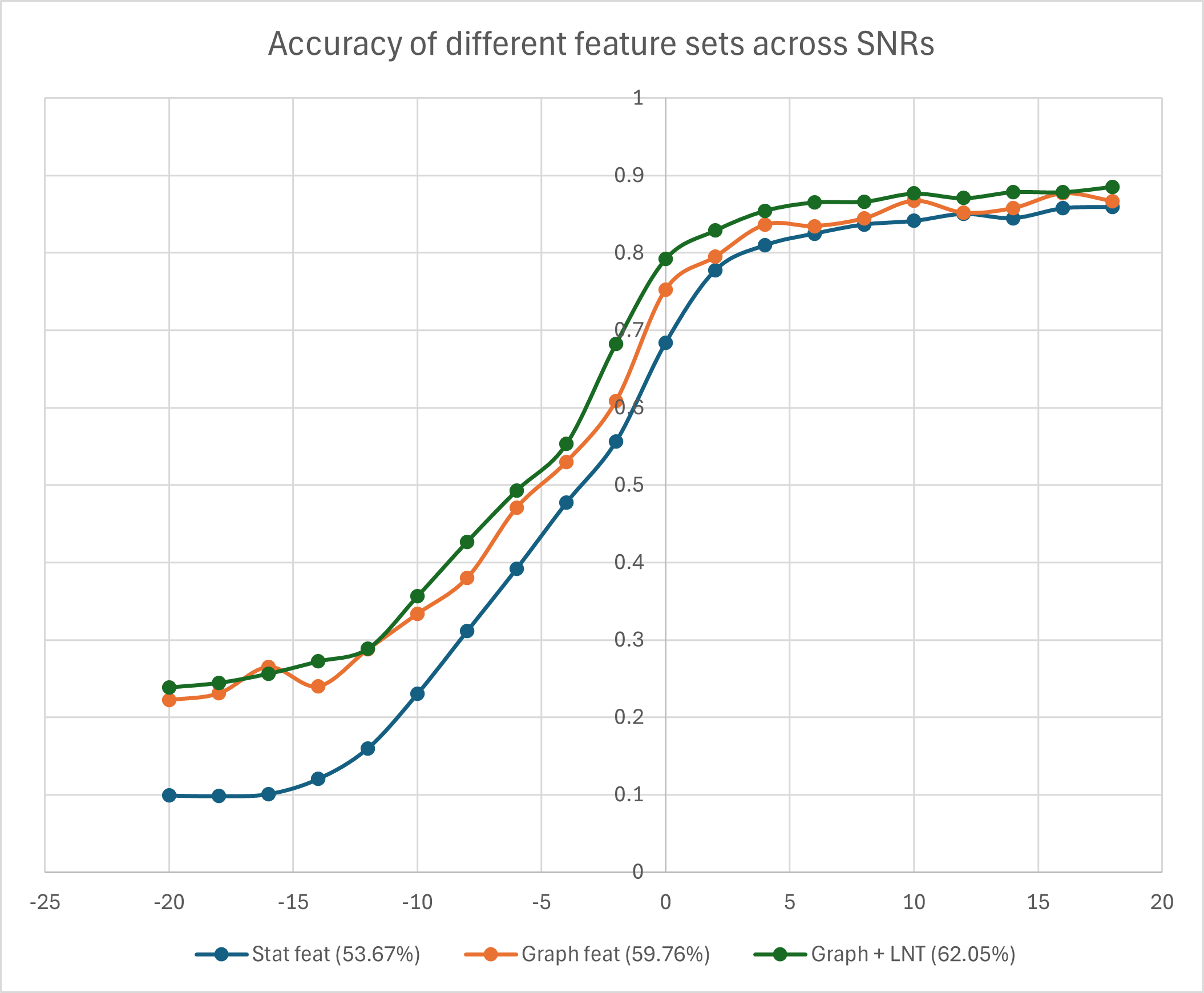}
\caption{accuracy of different feature sets across different SNRs}
\label{feats_acc}
\end{figure}

\begin{figure}[htbp]  
\centering
\includegraphics[width=1\columnwidth]{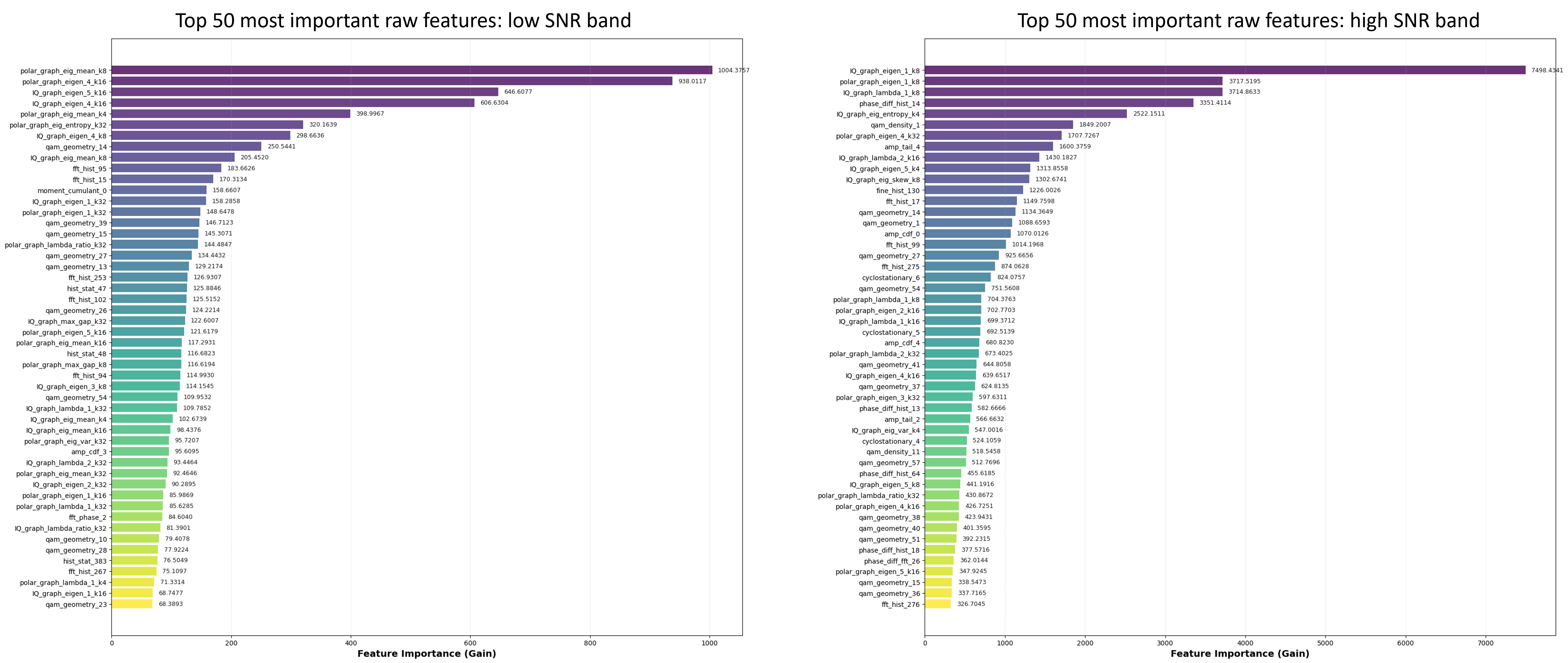}
\caption{Feature Importance Analysis of the raw feature set for the low (left) and high (right) SNR classifier}
\label{feat_import_raw}
\end{figure}

To evaluate the proposed spatio-temporal graph features across varying SNRs, we constructed multi-scale representations using $k \in \{4, 8, 16, 32\}$ nearest neighbors and a temporal penalty of $\lambda = 50$ to balance spatial geometry with temporal progression. As shown in Fig. \ref{feats_acc}, the graph features alone clearly outperform the baseline statistical features, improving the overall accuracy from 53.67\% to 59.76\%. This gap is especially prominent in the highly challenging -20 dB to 0 dB regime, demonstrating that topological connectivity is inherently more robust to severe noise than traditional density- or moment-based statistics. Feature importance analysis further validates this robustness. The model evaluates the utility of the feature using Information Gain \cite{chen2016xgboost}, which measures the reduction in the objective loss function after a split of the tree nodes:
\begin{equation}
  \text{Gain} = \frac{1}{2} \left[ \frac{G_L^2}{H_L + \lambda} + \frac{G_R^2}{H_R + \lambda} - \frac{(G_L + G_R)^2}{H_L + H_R + \lambda} \right] - \gamma 
\end{equation}
where $G$ and $H$ are the sums of the first and second-order gradients of the loss function for the left ($L$) and right ($R$) child nodes, $\lambda$ is the L2 regularization term, and $\gamma$ is the minimum loss reduction required for a split. Using the metric in Fig. \ref{feat_import_raw}, graph-derived spectral features (e.g., leading eigenvalues and spectral entropy) completely dominate the low SNR band, suppressing traditional statistical moments. This shows that under severe amplitude and phase corruption, macroscopic topological invariants provide the most reliable discriminative signal. Conversely, in the high-SNR band, fine-grained traditional metrics (such as phase difference histograms) re-emerge alongside graph eigenvalues. This complementary behavior confirms our dual-feature approach: graph features anchor baseline reliability in extreme noise, while traditional statistics synergize to capture microscopic constellation differences in clear channels.

\subsubsection{Linear Projection Features}

\begin{figure}[htbp]  
\centering
\includegraphics[width=1\columnwidth]{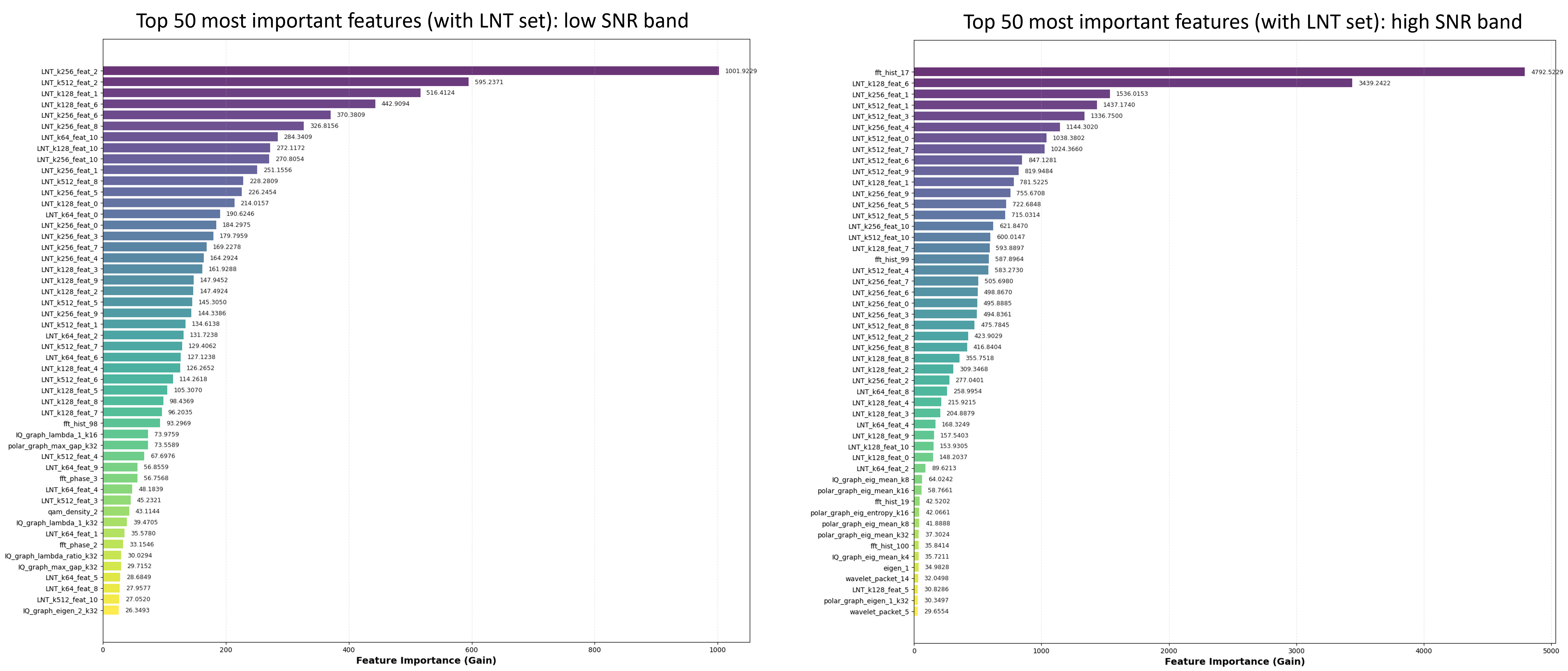}
\caption{Feature Importance Analysis of the learning feature set for the low (left) and high (right) SNR classifier}
\label{feats_lnt}
\end{figure}

Furthermore, to fully exploit the synergy between the spatio-temporal graph and statistical features, the combined feature set is evaluated by an auxiliary XGBoost model (configured identically to the downstream Mixture of Experts classifiers) for targeted dimensionality reduction. Based on the calculated Information Gain values, we extract multiple feature subspaces by isolating the 64, 128, 256, and 512 most important features. These multi-resolution subsets are independently processed through the Least-squares Normal Transform (LNT) module. To ensure robust, leak-free representations, the LNT employs a 5-fold cross-validation strategy, selecting the projection matrix from the best-performing fold to generate the final transformed features. As shown by the green line in Fig. \ref{feats_lnt}, this multi-subspace projection improves the system's overall accuracy to 61.29\%. Moreover, subsequent feature importance analysis reveals that these multidimensional LNT features dominate the highest rankings across both low- and high-SNR bands, confirming that this targeted projection successfully extracts complementary, highly discriminative signatures that significantly enrich the overall feature space.

\subsubsection{Numbers of MoEs}

\begin{figure}[htbp] 
\centering
\includegraphics[width=1\columnwidth]{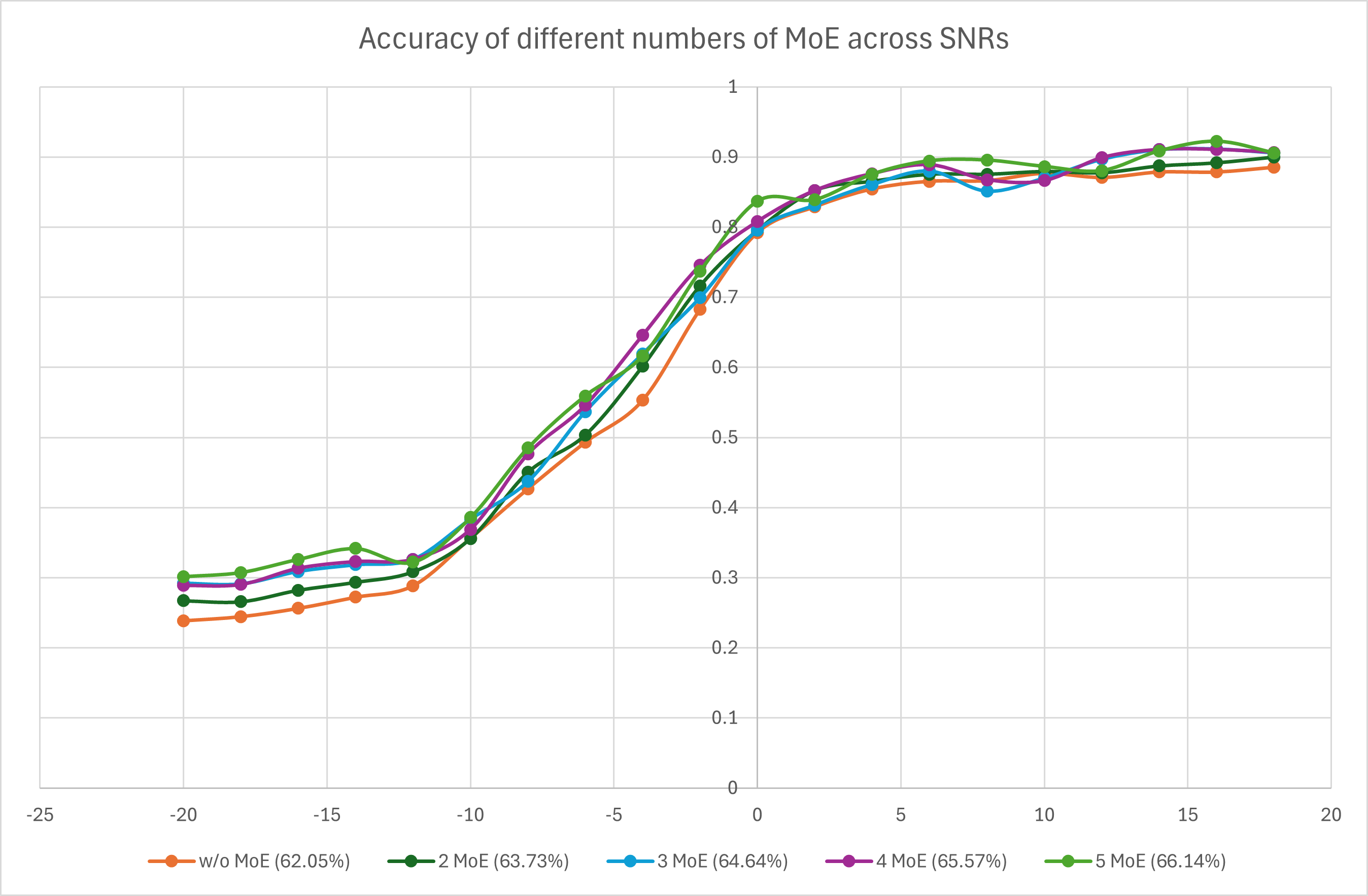}
\caption{accuracy of different numbers of MoE across SNRs}
\label{moe}
\end{figure}

To evaluate the effectiveness of the proposed Signal-to-Noise Ratio (SNR) soft-routing mechanism, we analyze classification accuracy across varying SNR levels using different Mixture of Experts (MoE) configurations. As shown in the performance curves in Fig. \ref{moe}, employing a single global classifier without routing ("w/o MoE", orange line) yields an overall accuracy of 61.29\%. By partitioning the dataset based on channel quality and routing samples to specialized expert classifiers, the system shows consistent performance gains.

\begin{figure*}[htbp]  
\centering
\includegraphics[width=1\textwidth]{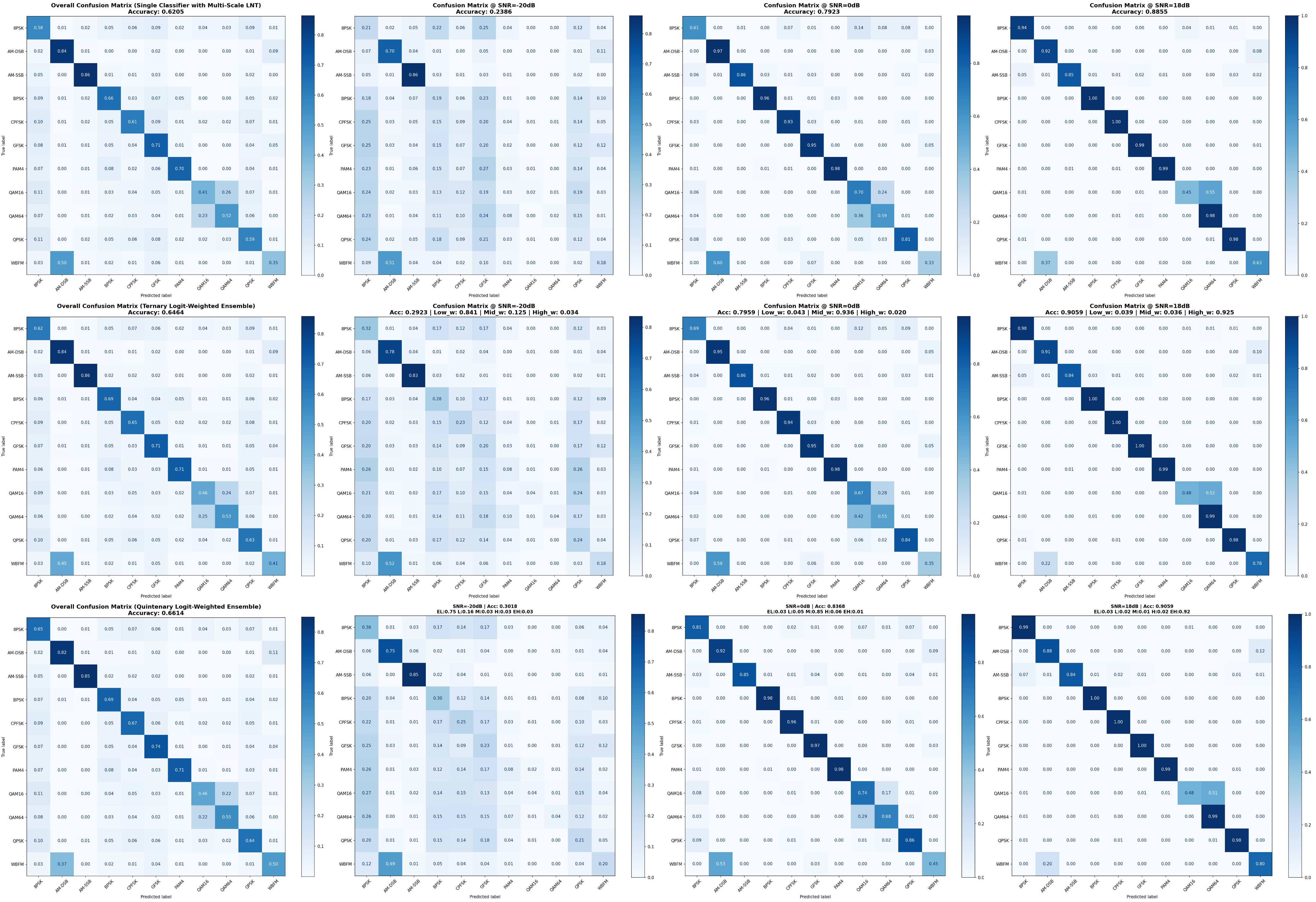}
\caption{Confusion Matrices at representative SNRs. The prediction accuracy of w/o MoE, 3-MoEs, and 5-MoEs is from top to bottom. The overall accuracy, -20dB, 0dB, and 18dB are from left to right.}
\label{conf}
\end{figure*}

Increasing the number of experts steadily improves overall accuracy: the 2 MoE, 3 MoE, and 4 MoE configurations achieve 63.73\%, 64.64\%, and 65.57\%, respectively. The proposed 5 MoE architecture (light green line) achieves an overall accuracy of 66.14\%. The advantage of finer SNR partitioning is particularly pronounced in the extreme low-SNR regime (-20 dB to -5 dB). By dedicating specific XGBoost experts to narrow, highly corrupted noise profiles, higher-order MoE architectures effectively isolate and mitigate severe feature-space domain shifts caused by volatile channel conditions. The confusion matrices for representative SNRs and the average accuracy are shown in Fig. \ref{conf}.

It is worth noting that while the MoE configurations heavily outperform the baseline on average, there are minor localized dips in performance in the mid-to-high SNR range (e.g., the 3 MoE briefly drops below the baseline at 8 dB). These fluctuations are characteristic of boundary transition effects; at thresholds between expert models, routing uncertainty and overlapping feature distributions can temporarily affect classification. Nevertheless, the five MoE configuration recovers rapidly and maintains the highest stability across the distribution tails, making it the most robust framework overall.

\begin{table}[htbp]
\caption{XGBoost Hyperparameters for Each MoE Expert Classifier}
\label{tab:moe-hyperparameters}
\centering
\begin{tabular}{|l|c|}
\hline
\textbf{Parameter} & \textbf{Value} \\
\hline
learning\_rate & 0.1 \\
max\_depth & 2 \\
n\_estimators & 200 \\
subsample & 0.8 \\
colsample\_bytree & 0.8 \\
min\_child\_weight & 3 \\
gamma & 0.1 \\
tree\_method & hist \\
\hline
\end{tabular}
\end{table}

\begin{table}[htbp]
\centering
\caption{Classification accuracy (\%) under different SNR ranges.}
\label{tab:low_snr_accuracy}
\resizebox{\columnwidth}{!}{
\begin{tabular}{lccc}
\hline
\textbf{Method} & \textbf{$-20 \sim -2$ dB} & \textbf{$0 \sim 18$ dB} & \textbf{$-20 \sim 18$ dB} \\
\hline
MCL-DNN \cite{xu2020spatiotemporal} & 33.29 & 91.06 & 62.12 \\
PET-CGDNN \cite{zhang2021efficient} & 32.55 & 90.84 & 61.65 \\
LENet-M \cite{wei2023family} & 35.87 & \textbf{93.44} & 64.63 \\
KADNet \cite{wang2026kadnet} & 36.86 & 92.45 & 64.66 \\
\hline
GAMC 5-MoE (Ours) & \textbf{43.41} & 88.8 & \textbf{66.14} \\
\end{tabular}
}
\end{table}

\subsection{Model Size and Complexity Analysis}

Table \ref{tab:complexity_bd} presents a detailed breakdown of the model parameters and floating-point operations (FLOPs) across the core architectural components. A key observation is that the Feature Extraction module is strictly mathematical and parameter-free (0 K). However, it establishes the baseline computational cost by demanding a constant of 2,796.2 K FLOPs across all configurations. As the system scales from the base GAMC model to the 5-MoE, the parameters and FLOPs associated with the Least-squares Normal Transform (LNT) and Mixture of Experts (MoE) modules increase linearly to support the additional routing branches. However, because the computationally dominant feature extraction phase is shared globally, the penalty for adding these experts is extremely minimal. The total number of parameters expands from 33.52 K to 177.6 K. In comparison, the total FLOPs increase only marginally from 2,823.92 K to 2,938.2 K. This confirms that the architecture effectively isolates the heavy processing to a single shared frontend, allowing the backend to execute highly complex, multi-expert routing with almost negligible overhead, perfectly aligning with the tight constraints of green edge-computing devices.

\begin{table}[htbp]
\centering
\caption{Model Parameters (K) and FLOPs (K) Analysis}
\label{tab:complexity_bd}
\resizebox{\columnwidth}{!}{
\begin{tabular}{lcccc}
\hline
Model & Feature Extraction & LNT & MoE & Total \\
\hline
GAMC w/o MoE & (0, 2,796.2) & (11.52, 21.12) & (22, 6.6) & (33.52, 2,823.92) \\

GAMC 2-MoE & (0, 2,796.2) & (23.04, 42.4) & (44, 13.2) & (69.04, 2,853.8) \\

GAMC 3-MoE & (0, 2,796.2) & (34.56, 63.6) & (66, 19.8) & (106.56, 2,885.6) \\

GAMC 4-MoE & (0, 2,796.2) & (46.08, 84.8) & (88, 26.4) & (142.08, 2,909.8) \\

GAMC 5-MoE & (0, 2,796.2) & (57.6, 106) & (110, 33) & (177.6, 2,938.2) \\

\hline
\end{tabular}
}
\end{table}

\begin{table*}[htbp]
\caption{Comparison to state-of-the-art methods on RadioML 2016.10A.}
\label{tab:comparison-mobilera}
\centering
\begin{tabular}{lccc}
\hline
\textbf{Methods} & \textbf{AMC Accuracy (\%)} & \textbf{Number of Weights (K)}  & \textbf{FLOPs (M)}\\
\hline
ResNet\cite{o2018over}  & 57.8 & 241.8 ($1.36\times$)& 11.49 ($3.91\times$)\\ 
LSTM \cite{rajendran2018deep} & 60.3 & 201.1 ($1.13\times$)& 25.82 ($8.78\times$)\\
MCLDNN\cite{xu2020spatiotemporal} & 62.1 & 406.1 ($2.28\times$)& 35.8 ($12.18\times$)\\ 
PET-CGDNN\cite{zhang2021efficient} & 61.7 & 71.87 ($0.40\times$)& 4.6 ($1.56\times$)\\ 
MobileAmcT\cite{fei2024mobileamct} & 62.9 & 575.06 ($3.23\times$)& 8.90 ($3.03\times$)\\
AbFTNet\cite{ning2024abftnet} & 64.6 & 175.69 ($0.99\times$)& 6.96 ($2.37\times$)\\
LENet-M\cite{wei2023family} & 64.6 & 55.86 ($0.31\times$)& 1.9 ($0.65\times$)\\
TDRNN\cite{an2025efficient} & 63.5 & 41.82 ($0.23\times$)& 10.96 ($3.73\times$)\\ 
KADNet\cite{wang2026kadnet} & 64.7 & 345 ($1.94\times$)& 12.52 ($4.26\times$)\\
AvgNet\cite{xuan2022avgnet} & 62.9 & 1,830 ($10.28\times$)& - \\ 
GIGNet\cite{ke2025gignet} & 63.8 & 2,550 ($14.32\times$)& - \\

MobileRaT-B\cite{zheng2023mobilerat} ($1\times$) & 65.9 & 268 ($1.51\times$)& 92.31 ($31.40\times$)\\
MobileRaT-B\cite{zheng2023mobilerat} ($0.38\times$) & 63.2 & 102 ($0.57\times$) & 35.15 ($11.96\times$)\\
\hline
GAMC w/o MoE (ours)   & 62.1 & \textbf{34} ($0.19\times$) & \textbf{2.82} ($0.96\times$)\\
GAMC 2-MoE (ours) & 63.7 & 69 ($0.39\times$) & 2.85 ($0.97\times$)\\
GAMC 3-MoE (ours) & 64.6 & 107 ($0.60\times$)& 2.88 ($0.98\times$)\\
GAMC 4-MoE (ours) & 65.6 & 142 ($0.80\times$)& 2.91 ($0.99\times$)\\
GAMC 5-MoE (ours) & \textbf{66.1} & 178 ($1\times$)& 2.94 ($1\times$)\\
\end{tabular}
\end{table*}

\subsection{Comparison with State-of-the-Art Methods}

To comprehensively evaluate the proposed architecture, we benchmarked GAMC against a diverse array of state-of-the-art methods on the RadioML 2016.10A dataset \cite{wz93-h307-25}. Our baselines span several distinct design paradigms: CNN-based spatial models (ResNet \cite{o2018over}), RNN-based temporal architectures (LSTM \cite{rajendran2018deep}), and spatiotemporal fusion networks (MCLDNN \cite{xu2020spatiotemporal}). We also included recent mobile Transformer frameworks (MobileAmcT \cite{fei2024mobileamct}, MobileRaT-B \cite{zheng2023mobilerat}) and graph-based methodologies (AvgNet \cite{xuan2022avgnet}, GIGNet \cite{ke2025gignet}). Finally, to rigorously assess computational complexity and noise resilience, we compared our approach with lightweight efficiency-focused models (LENet-M \cite{wei2023family}, TDRNN \cite{an2025efficient}, PET-CGDNN \cite{zhang2021efficient}, AbFTNet \cite{ning2024abftnet}) as well as KADNet \cite{wang2026kadnet}, a model explicitly specialized for low-SNR conditions. As detailed in Table V, the GAMC 5-MoE configuration achieves a maximum overall accuracy of 66.1\%, outperforming the classification performance of the relatively computationally heavy, transformer-based MobileRaT-B model (1$\times$). Crucially, our proposed method achieves this while maintaining an interpretability and lightweight profile, requiring only 178K trainable parameters and 2.94M FLOPs. This represents an approximate 31-time reduction in computational complexity compared to MobileRaT-B's 92.31M FLOPs. Furthermore, Table IV highlights the architecture's exceptional resilience to severe channel degradation. In Table \ref{tab:low_snr_accuracy}, we compared GAMC with other methods in the low-SNR regime (-20 to -2 dB), and the high-SNR regime (0 to 18 dB). In the highly challenging low-SNR regime, GAMC 5-MoE achieves a 43.41\% accuracy, outperforming the closest baseline, KADNet (36.86\%), by a substantial margin of over 6.5\%. Although some deep learning models marginally exceed our performance in perfectly clean, high-SNR conditions, our method establishes a vastly superior holistic balance. By combining macroscopic topological graph features with dynamic soft-routing, GAMC delivers noise robustness at a fraction of the computational costs of traditional approaches, making it well-suited to resource-constrained edge deployment.

\section{Conclusion}
In this work, a novel, lightweight, learning-based automatic modulation classifier, called GAMC, was proposed. By fusing robust statistical characteristics with dual-manifold spatio-temporal graph features and subsequently refining them via a multi-subspace LNT projection, GAMC extracts highly discriminative, noise-resilient signal representations. To mitigate feature-space domain shifts caused by fluctuating channel conditions, GAMC adopts a Mixture of Experts (MoE) mechanism guided by a lightweight Channel Quality Indicator (CQI). Experimental evaluations on the RadioML 2016.10a dataset demonstrated the superiority of GAMC. Its 5-MoE configuration achieves an overall competitive accuracy of 66.14\%. It also exhibits resilience to severe channel degradation, achieving 43.41\% accuracy in the extreme low-SNR regime (-20 to -2 dB), significantly outperforming existing baselines. GAMC accomplishes this with a low computational footprint, requiring only 177.6K trainable parameters and 2.94M FLOPs. By delivering uncompromised noise robustness at a fraction of the computational cost of traditional deep learning models, GAMC establishes an optimal trade-off between classification accuracy and model complexity, making it a practical solution for real-time deployment on resource-constrained edge AI devices.


\bibliographystyle{IEEEtran}
\bibliography{refs}

@article{wei2019automatic,
  title={Automatic modulation classification of digital communication signals using SVM based on hybrid features, cyclostationary, and information entropy},
  author={Wei, Yangjie and Fang, Shiliang and Wang, Xiaoyan},
  journal={Entropy},
  volume={21},
  number={8},
  pages={745},
  year={2019},
  publisher={MDPI}
}

@article{zhang2019automatic,
  title={Automatic modulation classification using convolutional neural network with features fusion of SPWVD and BJD},
  author={Zhang, Zufan and Wang, Chun and Gan, Chenquan and Sun, Shaohui and Wang, Mengjun},
  journal={IEEE Transactions on Signal and Information Processing over Networks},
  volume={5},
  number={3},
  pages={469--478},
  year={2019},
  publisher={IEEE}
}

@inproceedings{wang2017graphic,
  title={Graphic constellations and DBN based automatic modulation classification},
  author={Wang, Fen and Wang, Yongchao and Chen, Xi},
  booktitle={2017 IEEE 85th vehicular technology conference (VTC Spring)},
  pages={1--5},
  year={2017},
  organization={IEEE}
}

@article{zhu2014genetic,
  title={Genetic algorithm optimized distribution sampling test for M-QAM modulation classification},
  author={Zhu, Zhechen and Aslam, Muhammad Waqar and Nandi, Asoke K},
  journal={Signal Processing},
  volume={94},
  pages={264--277},
  year={2014},
  publisher={Elsevier}
}

@article{wang2010fast,
  title={Fast and robust modulation classification via Kolmogorov-Smirnov test},
  author={Wang, Fanggang and Wang, Xiaodong},
  journal={IEEE Transactions on Communications},
  volume={58},
  number={8},
  pages={2324--2332},
  year={2010},
  publisher={IEEE}
}

@article{chang2021multitask,
  title={Multitask-learning-based deep neural network for automatic modulation classification},
  author={Chang, Shuo and Huang, Sai and Zhang, Ruiyun and Feng, Zhiyong and Liu, Liang},
  journal={IEEE internet of things journal},
  volume={9},
  number={3},
  pages={2192--2206},
  year={2021},
  publisher={IEEE}
}

@inproceedings{hamidi2021mcformer,
  title={Mcformer: A transformer based deep neural network for automatic modulation classification},
  author={Hamidi-Rad, Shahab and Jain, Swayambhoo},
  booktitle={2021 IEEE Global Communications Conference (GLOBECOM)},
  pages={1--6},
  year={2021},
  organization={IEEE}
}

@article{kim2021deep,
  title={Deep learning-based robust automatic modulation classification for cognitive radio networks},
  author={Kim, Seung-Hwan and Kim, Jae-Woo and Nwadiugwu, Williams-Paul and Kim, Dong-Seong},
  journal={IEEE access},
  volume={9},
  pages={92386--92393},
  year={2021},
  publisher={IEEE}
}

@article{ramkumar2009automatic,
  title={Automatic modulation classification for cognitive radios using cyclic feature detection},
  author={Ramkumar, Barathram},
  journal={IEEE Circuits and Systems Magazine},
  volume={9},
  number={2},
  pages={27--45},
  year={2009},
  publisher={IEEE}
}

@article{ge2021accuracy,
  title={Accuracy analysis of feature-based automatic modulation classification via deep neural network},
  author={Ge, Zhan and Jiang, Hongyu and Guo, Youwei and Zhou, Jie},
  journal={Sensors},
  volume={21},
  number={24},
  pages={8252},
  year={2021},
  publisher={MDPI}
}

@article{wei2000maximum,
  title={Maximum-likelihood classification for digital amplitude-phase modulations},
  author={Wei, Wen and Mendel, Jerry M},
  journal={IEEE transactions on Communications},
  volume={48},
  number={2},
  pages={189--193},
  year={2000},
  publisher={IEEE}
}

@inproceedings{sills1999maximum,
  title={Maximum-likelihood modulation classification for PSK/QAM},
  author={Sills, James A},
  booktitle={MILCOM 1999. IEEE Military Communications. Conference Proceedings (Cat. No. 99CH36341)},
  volume={1},
  pages={217--220},
  year={1999},
  organization={IEEE}
}

@article{abu2018automatic,
  title={Automatic modulation classification using moments and likelihood maximization},
  author={Abu-Romoh, Mohannad and Aboutaleb, Ahmed and Rezki, Zouheir},
  journal={IEEE Communications Letters},
  volume={22},
  number={5},
  pages={938--941},
  year={2018},
  publisher={IEEE}
}

@inproceedings{ghasemzadeh2018performance,
  title={Performance evaluation of feature-based automatic modulation classification},
  author={Ghasemzadeh, Pejman and Banerjee, Subharthi and Hempel, Michael and Sharif, Hamid},
  booktitle={2018 12th international conference on signal processing and communication systems (ICSPCS)},
  pages={1--5},
  year={2018},
  organization={IEEE}
}

@article{wu2008novel,
  title={Novel automatic modulation classification using cumulant features for communications via multipath channels},
  author={Wu, Hsiao-Chun and Saquib, Mohammad and Yun, Zhifeng},
  journal={IEEE Transactions on Wireless Communications},
  volume={7},
  number={8},
  pages={3098--3105},
  year={2008},
  publisher={IEEE}
}

@article{o2018over,
  title={Over-the-air deep learning based radio signal classification},
  author={O’Shea, Timothy James and Roy, Tamoghna and Clancy, T Charles},
  journal={IEEE Journal of Selected Topics in Signal Processing},
  volume={12},
  number={1},
  pages={168--179},
  year={2018},
  publisher={IEEE}
}

@article{zheng2023mobilerat,
  title={MobileRaT: a lightweight radio transformer method for automatic modulation classification in drone communication systems},
  author={Zheng, Qinghe and Tian, Xinyu and Yu, Zhiguo and Ding, Yao and Elhanashi, Abdussalam and Saponara, Sergio and Kpalma, Kidiyo},
  journal={Drones},
  volume={7},
  number={10},
  pages={596},
  year={2023},
  publisher={MDPI}
}

@article{zheng2025recent,
  title={Recent advances in automatic modulation classification technology: Methods, results, and prospects},
  author={Zheng, Qinghe and Tian, Xinyu and Yu, Lisu and Elhanashi, Abdussalam and Saponara, Sergio},
  journal={International Journal of Intelligent Systems},
  volume={2025},
  number={1},
  pages={4067323},
  year={2025},
  publisher={Wiley Online Library}
}

@article{harper2023automatic,
  title={Automatic modulation classification with deep neural networks},
  author={Harper, Clayton A and Thornton, Mitchell A and Larson, Eric C},
  journal={Electronics},
  volume={12},
  number={18},
  pages={3962},
  year={2023},
  publisher={MDPI}
}

@article{rajendran2018deep,
  title={Deep learning models for wireless signal classification with distributed low-cost spectrum sensors},
  author={Rajendran, Sreeraj and Meert, Wannes and Giustiniano, Domenico and Lenders, Vincent and Pollin, Sofie},
  journal={IEEE Transactions on Cognitive Communications and Networking},
  volume={4},
  number={3},
  pages={433--445},
  year={2018},
  publisher={IEEE}
}

@article{zhang2017modulation,
  title={Modulation classification in multipath fading channels using sixth-order cumulants and stacked convolutional auto-encoders},
  author={Zhang, Zhenyu and Hua, Zhong and Liu, Yingzhe},
  journal={IET communications},
  volume={11},
  number={6},
  pages={910--915},
  year={2017},
  publisher={Wiley Online Library}
}

@inproceedings{liu2017deep,
  title={Deep neural network architectures for modulation classification},
  author={Liu, Xiaoyu and Yang, Diyu and El Gamal, Aly},
  booktitle={2017 51st Asilomar Conference on Signals, Systems, and Computers},
  pages={915--919},
  year={2017},
  organization={IEEE}
}

@inproceedings{pijackova2021radio,
  title={Radio modulation classification using deep learning architectures},
  author={Pijackova, Kristyna and Gotthans, Tomas},
  booktitle={2021 31st international conference radioelektronika (radioelektronika)},
  pages={1--5},
  year={2021},
  organization={IEEE}
}

@article{wei2020intra,
  title={Intra-pulse modulation radar signal recognition based on CLDN network},
  author={Wei, Shunjun and Qu, Qizhe and Su, Hao and Wang, Mou and Shi, Jun and Hao, Xiaojun},
  journal={IET Radar, Sonar \& Navigation},
  volume={14},
  number={6},
  pages={803--810},
  year={2020},
  publisher={Wiley Online Library}
}

@inproceedings{wu2022deep,
  title={Deep multi-scale representation learning with attention for automatic modulation classification},
  author={Wu, Xiaowei and Wei, Shengyun and Zhou, Yan},
  booktitle={2022 International Joint Conference on Neural Networks (IJCNN)},
  pages={1--8},
  year={2022},
  organization={IEEE}
}

@article{zheng2021spectrum,
  title={Spectrum interference-based two-level data augmentation method in deep learning for automatic modulation classification},
  author={Zheng, Qinghe and Zhao, Penghui and Li, Yang and Wang, Hongjun and Yang, Yang},
  journal={Neural Computing and Applications},
  volume={33},
  number={13},
  pages={7723--7745},
  year={2021},
  publisher={Springer}
}

@article{zheng2023dl,
  title={DL-PR: Generalized automatic modulation classification method based on deep learning with priori regularization},
  author={Zheng, Qinghe and Tian, Xinyu and Yu, Zhiguo and Wang, Hongjun and Elhanashi, Abdussalam and Saponara, Sergio},
  journal={Engineering Applications of Artificial Intelligence},
  volume={122},
  pages={106082},
  year={2023},
  publisher={Elsevier}
}

@article{an2025efficient,
  title={Efficient automatic modulation classification for next-generation wireless networks},
  author={An, To Truong and Argyriou, Antonios and Puspitasari, Annisa Anggun and Cotton, Simon L and Lee, Byung Moo},
  journal={IEEE Transactions on Green Communications and Networking},
  year={2025},
  publisher={IEEE}
}

@article{wang2026kadnet,
  title={KADNet: Low SNR Automatic Modulation Classification via SNR Aware Deformable Convolution and Kolmogorov--Arnold Networks},
  author={Wang, Run and Li, Jizhe and Yang, Youze and Wang, Shasha and Zheng, Bing},
  journal={Digital Signal Processing},
  pages={105942},
  year={2026},
  publisher={Elsevier}
}

@article{tahir2026hfdnn,
  title={HFDNN: A Hybrid Fusion Deep Neural Network for Robust Automatic Modulation Classification in Adverse Wireless Environments},
  author={Tahir, M Muneeb and Latif, Arbab and Younis, M Shahzad and Rais, Rao Naveed Bin and Ammar, Khalid},
  journal={IEEE Access},
  year={2026},
  publisher={IEEE}
}

@article{zheng2022fine,
  title={Fine-grained modulation classification using multi-scale radio transformer with dual-channel representation},
  author={Zheng, Qinghe and Zhao, Penghui and Wang, Hongjun and Elhanashi, Abdussalam and Saponara, Sergio},
  journal={IEEE Communications Letters},
  volume={26},
  number={6},
  pages={1298--1302},
  year={2022},
  publisher={IEEE}
}

@article{ahmadi2025enhancing,
  title={Enhancing automatic modulation recognition with a reconstruction-driven vision transformer under limited labels},
  author={Ahmadi, Hossein and Saffari, Banafsheh and Mahdimahalleh, Sajjad Emdadi and Safari, Mohammad Esmaeil and Ahmadi, Aria},
  journal={arXiv preprint arXiv:2508.20193},
  year={2025}
}

@article{gao2023moe,
  title={Moe-amc: Enhancing automatic modulation classification performance using mixture-of-experts},
  author={Gao, Jiaxin and Cao, Qinglong and Chen, Yuntian},
  journal={arXiv preprint arXiv:2312.02298},
  year={2023}
}

@article{meng2025adversarial,
  title={Adversarial attack and reliable defense based on frequency domain feature enhancement for automatic modulation classification},
  author={Meng, Yongchao and Qi, Peihan and Zheng, Shilian and Cai, Zihao and Zhou, Xiaoyu and Jiang, Tao},
  journal={IEEE Transactions on Information Forensics and Security},
  year={2025},
  publisher={IEEE}
}

@article{ansari2025attention,
  title={Attention-enhanced hybrid automatic modulation classification for advanced wireless communication systems: A deep learning-transformer framework},
  author={Ansari, Sam and Alnajjar, Khawla A and Majzoub, Sohaib and Almajali, Eqab and Jarndal, Anwar and Bonny, Talal and Hussain, Abir and Mahmoud, Soliman},
  journal={IEEE Access},
  year={2025},
  publisher={IEEE}
}

@article{xuan2022avgnet,
  title={AvgNet: Adaptive visibility graph neural network and its application in modulation classification},
  author={Xuan, Qi and Zhou, Jinchao and Qiu, Kunfeng and Chen, Zhuangzhi and Xu, Dongwei and Zheng, Shilian and Yang, Xiaoniu},
  journal={IEEE Transactions on Network Science and Engineering},
  volume={9},
  number={3},
  pages={1516--1526},
  year={2022},
  publisher={IEEE}
}

@article{wang2023automatic,
  title={Automatic modulation classification based on CNN-transformer graph neural network},
  author={Wang, Dong and Lin, Meiyan and Zhang, Xiaoxu and Huang, Yonghui and Zhu, Yan},
  journal={Sensors},
  volume={23},
  number={16},
  pages={7281},
  year={2023},
  publisher={MDPI}
}

@article{xu2020spatiotemporal,
  title={A spatiotemporal multi-channel learning framework for automatic modulation recognition},
  author={Xu, Jialang and Luo, Chunbo and Parr, Gerard and Luo, Yang},
  journal={IEEE Wireless Communications Letters},
  volume={9},
  number={10},
  pages={1629--1632},
  year={2020},
  publisher={IEEE}
}

@article{zhang2021efficient,
  title={An efficient deep learning model for automatic modulation recognition based on parameter estimation and transformation},
  author={Zhang, Fuxin and Luo, Chunbo and Xu, Jialang and Luo, Yang},
  journal={IEEE Communications Letters},
  volume={25},
  number={10},
  pages={3287--3290},
  year={2021},
  publisher={IEEE}
}

@article{ke2025gignet,
  title={GIGNet: A graph-in-graph neural network for automatic modulation recognition},
  author={Ke, Yang and Zhang, Wancheng and Zhang, Yan and Zhao, Haoyu and Fei, Zesong},
  journal={IEEE Transactions on Vehicular Technology},
  year={2025},
  publisher={IEEE}
}

@article{fei2024mobileamct,
  title={Mobileamct: A lightweight mobile automatic modulation classification transformer in drone communication systems},
  author={Fei, Hongyun and Wang, Baiyang and Wang, Hongjun and Fang, Ming and Wang, Na and Ran, Xingping and Liu, Yunxia and Qi, Min},
  journal={Drones},
  volume={8},
  number={8},
  pages={357},
  year={2024},
  publisher={MDPI}
}

@article{ning2024abftnet,
  title={Abftnet: An efficient transformer network with alignment before fusion for multimodal automatic modulation recognition},
  author={Ning, Meng and Zhou, Fan and Wang, Wei and Wang, Shaoqiang and Zhang, Peiying and Wang, Jian},
  journal={Electronics},
  volume={13},
  number={18},
  pages={3725},
  year={2024},
  publisher={MDPI}
}

@article{wei2023family,
  title={A family of automatic modulation classification models based on domain knowledge for various platforms},
  author={Wei, Shengyun and Wang, Zhenyi and Sun, Zhaolong and Liao, Feifan and Li, Zhen and Zou, Li and Mi, Haibo},
  journal={Electronics},
  volume={12},
  number={8},
  pages={1820},
  year={2023},
  publisher={MDPI}
}

@inproceedings{wang2024statistics,
  title={A statistics-based feature generation (sfg) method: Theory and applications},
  author={Wang, Xinyu and Wu, Yixing and Li, Haiyi and Mishra, Vinod K and Kuo, C-C Jay},
  booktitle={2024 IEEE International Conference on Big Data (BigData)},
  pages={5731--5738},
  year={2024},
  organization={IEEE}
}

@misc{wz93-h307-25,
doi = {10.21227/wz93-h307},
url = {https://dx.doi.org/10.21227/wz93-h307},
author = {T.J. O'Shea},
publisher = {IEEE Dataport},
title = {RadioML2016.10A/10B/10C},
year = {2025} }

@inproceedings{chen2016xgboost,
  title={Xgboost: A scalable tree boosting system},
  author={Chen, Tianqi and Guestrin, Carlos},
  booktitle={Proceedings of the 22nd acm sigkdd international conference on knowledge discovery and data mining},
  pages={785--794},
  year={2016}
}

@misc{yu2025gamc,
  title        = {G-AMC: A Green Automatic Modulation Classification Method},
  author       = {Yu, Chee-An and Chen, Young-Kai and Kuo, C.-C. Jay},
  year         = {2025},
  note         = {Accepted to IEEE GLOBECOM 2025 Workshop},
  eprint       = {2604.06402},
  archivePrefix= {arXiv},
  primaryClass = {eess.SP}
}

\end{document}